# ExoMars Raman Laser Spectrometer (RLS) – a tool for the potential recognition of wet-target craters on Mars

*Marco Veneranda[1], Guillermo Lopez-Reyes[1], José Antonio Manrique[1], Jesus Medina[1], Patricia Ruiz-Galende[2], Imanol Torre-Fdez[2], Kepa Castro[2], Cateline Lantz[3], Francois Poulet[3], Henning Dypvik[4], Stephanie C. Werner[4], Fernando Rull[1]*

[1] CSIC-CAB Associated Unit ERICA, Department of Condensed Matter Physics, University of Valladolid, Spain. Ave. Francisco Vallés, 8, Boecillo, 47151 Spain. marco.veneranda.87@gmail.com
[2] Department of Analytical Chemisty, University of the Basque Country (UPV/EHU), 48940 Leioa, Spain
[3] Institut d'Astrophysique Spatiale, CNRS/ Université Paris-Sud, France
[4] Department of Geosciences, CEED/GEO, University of Oslo, Norway

**ABSTRACT**

In the present work, NIR, LIBS, Raman and XRD techniques have been complementarily used to carry out a comprehensive characterization of a terrestrial analogue selected from the Chesapeake Bay Impact Structure (CBIS). The obtained data clearly highlight the key role of Raman spectroscopy in the detection of minor and trace compounds, through which inferences about geological processes occurred in the CBIS can be extrapolated. Beside the use of commercial systems, further Raman analyses were performed by the Raman Laser Spectrometer (RLS) ExoMars Simulator. This instrument represents the most reliable tool to effectively predict the scientific capabilities of the ExoMars/Raman system that will be deployed on Mars in 2021. By emulating the analytical procedures and operational restrictions established by the ExoMars mission rover design, it was proved that the RLS ExoMars Simulator is able to detect the amorphization of quartz, which constitutes an analytical clue of the impact origin of craters. On the other hand, the detection of barite and siderite, compounds crystallizing under hydrothermal conditions, helps to indirectly confirm the presence of water in impact targets. Furthermore, the RLS ExoMars Simulator capability of performing smart molecular mappings was also evaluated. According to the obtained results, the algorithms developed for its operation provide a great analytical advantage over most of the automatic analysis systems employed by commercial Raman instruments, encouraging its application for many additional scientific and commercial purposes.

**1 INTRODUCTION**

The analytical study of terrestrial analogues represents a cornerstone tool to deepen the knowledge about geological or biological processes occurred on extraterrestrial bodies (Werner *et al.*, 2018). From the study of terrestrial impact craters, many important information about the geology of Mars can be extrapolated (Spray *et al.*, 2010; Sturm *et al.*, 2013). Indeed, the morphological features of impact craters, being strongly affected by the composition and the properties of the target (Prieur *et al*., 2018) may provide supporting evidence about the past presence of liquid water on the Martian surface (Herrick and Hynek, 2017; Barlow and Perez, 2013). This is the case of the so called "inverted sombrero" craters (Horton Jr *et al*., 2006), which are characterized by a rimless annular trough extending far beyond the original transient crater. Following de Villiers *et al.* (2010), the presence of those peculiar craters on Earth is the consequence of the impact of extraterrestrial bodies into a "wet-target" (Ormo *et al.*, 2006) composed of three layers of different strength: (1) a water column, (2) a poorly consolidated lithological layer (i.e. water saturated sediments) and (3) a compact crystalline basement (i.e. continental crust).

The recognition of potential wet-target impact craters on Earth, which size and shape are mainly dependent on bolide size, water depth at impact site and target lithology, can be validated by the analytical characterization of geological samples. Specifically, the identification of shock-induced metamorphisms recognizable in impact breccia provides an understanding of the pressure reached during the impact (Wünnemann *et al.*, 2016)), while the mineralogical characterization of the stratigraphic units composing the crater helps to understand the characteristics of the target and its evolution (Robolledo-Vieyra and Urrutia-Fucugauchi, 2010).

According to the review of Dypvik and Jansa (2003) , less than 30 wet-target craters have been found on Earth, being the Chesapeake Bay Impact Structure (CBIS) widely recognized as the best preserved one (Gohn *et al.*, 2008). The study of this crater has great relevance in the field of planetary exploration since, through comparative examinations, they can help to identify the presence of wet-target impact structures on Mars. The study of Horton et al. (2006), based on the analysis of 531 images collected by the Mars Orbiter Camera wide-angle (MOC WA) over the Martian surface, provides a list of impact structures with a morphology comparable to the CBIS. Based on their interpretations, the extensive collapsed craters located in the Xanthe Terra and Margaritifer Terra are the most likely to be associated with wet-target impact structures. This is also consistent with the chaotic terrain of those areas assumed to be the consequence of the fluidization of ice water or clathrate hydrates from the subsurface (Rodriguez *et al.*, 2005).

Unlike craters on Earth, the recognition of impact structures on Mars currently based on their morphology only could not be validated by analytical data yet. However, the forthcoming ExoMars/ESA and Mars2020/NASA missions, to be launched in 2020, will deploy on Mars exploration rovers equipped with analytical tools that may provide the molecular and mineralogical data that are needed to confirm the impact origin of Martian craters among others (Lopez-Reyes *et al.*, 2013; Rull *et al.*, 2017).

The ExoMars rover will hold a suite of complementary analytical instruments (Pasteur Payload) dedicated to the characterization of geological samples collected from the Martian surface and sub-surface down to a depth of two meters (Debus *et al.*, 2010). Beside MicroOmega and MOMA instrument, the Pasteur payload includes the Raman Laser Spectrometer (RLS), which represents the first Raman system in history to be validated for space mission purposes. In this light, assessing the extent of the scientific discoveries that can be reached on Mars thorough Raman analysis has become of great importance. This is clearly corroborated by the increasing number of scientific studies dedicated to the Raman characterization of terrestrial analogue materials (Lalla *et al.*, 2016a; Lalla *et al.*, 2016b, Edwards *et al.*, 2007; Bost *et al.*, 2015).

Playing the leading role in the development of the ExoMars/RLS instrument (Rull *et al*., 2017), the UVa-CSIC-CAB Associated Unit ERICA (Spain) assembled the so called RLS ExoMars Simulator (Rull *et al*., 2017), a Raman spectrometer with analogue technical features to the system that will land on Mars. This analytical tool is coupled to a replicate of the Sample Preparation and Distribution System (SPDS) of the Pasteur suite and works in automatic mode by employing the same algorithms that the RLS will use on Mars for the adjustment of the acquisition parameters. Due to those unique features, the RLS ExoMars Simulator can be considered the most reliable tool to effectively emulate and predict the scientific capabilities of the RLS instrument on Mars.

Aiming at supporting the upcoming ExoMars/ESA mission, this work focuses on the geochemical and mineralogical characterization of a CBIS breccia sample thorough the complementary use of commercial analytical systems and the RLS ExoMars Simulator. In detail, the main aims of this research can be resumed as follow: 1) provide a comprehensive characterization of the selected terrestrial analogue using different analytical techniques, 2) deepen the knowledge about impact-related geological processes occurred in the CBIS and 3) evaluate the scientific capabilities of the ExoMars/RLS instrument in the discrimination of wet-target craters on Mars.

In a broader perspective, this study was performed in the framework of the Planetary Terrestrial Analogues Library (PTAL) project, which purpose is to build and provide the scientific community with a database of spectroscopic analyses obtained from a vast collection of terrestrial analogue samples selected due to their congruence with well-known mineralogical and geological features of extraterrestrial bodies.

## 2 MATERIALS AND METHODS

### 2.1 Description of the terrestrial analogue

*Chesapeake Bay Crater*

Seismic reflection data collected in the early 90s from southeastern Virginia (eastern shore of North America) enabled the detection of the CBIS, a complex impact structure buried between 200 and 500 meters beneath the surface (Powars *et al.,* 1993). The CBIS can be dated back to 35.3 million of years ago (Horton and Izett, 2005) and, according to numerical modeling performed by Collins and Wünnemann (2005), its shape (inverted sombrero) and total extension (85km) fit with the morphology of a crater produced by the impact of a bolide into a target covered by a thin layer of water.

Since 2000, numerous core holes have been drilled (Horton Jr. et al., 2005) to study the stratigraphic units of this impact structure. Among them, the ICDP-USGS Eyreville core, drilled between 2005 and 2006 to a total depth of 1766.3 m, provides the most complete lithologic section of the central crater (Horton Jr *et al.,* 2008).

To characterize the mineralogical composition of Eyreville Core samples, petrographical microscopes and X-ray diffractometers (XRD) (Horton Jr *et al.,* 2008; Bartosova *et al.,* 2010; Jackson *et al.,* 2011) have been extensively employed, while the use of Raman spectrometers is so far limited (Bartosova *et al.,* 2010; Jackson *et al.,* 2011; Jackson *et al.*, 2016). According to mineralogical data collected so far, the cored section consists of 6 main stratigraphic units (SU) (Rodriguez *et al.,* 2008).

*Sample selection*

For this work, the WH16-014 sample was selected from the upper melt-rich section of the suevitic and lithic impact breccias (5$^{th}$ SU), which represents the stratigraphic unit conserving the most clear impact-related mineralogical alterations. By definition, impact breccias are composed of a fine-grained matrix surrounding rock clasts of different composition and origin, in which a variable amount of amorphous materials (melts) and metamorphic effects induced by the impact can be detected. The upper part of the CBIS impact breccia (from 1397.2 to 1474.1 m) is mainly composed of suevites with 30% by volume of melt rocks. The lower part (from 1474.1 to 1551.2 m) shows a lower amount of melt particles and alternates layers of cataclastic gneiss blocks and polymict impact breccias (Rodriguez *et al.*, 2008).

The geological material selected for this work has a total weight of 21.4 g and was obtained from a depth of 1407.2 m. The sample belongs to a sub unit located just below the upper suevite, which is characterized by a melt matrix containing a large amount of mineral grains and lithic clasts. Previous XRD analysis detected quartz, sanidine, α-cristobalite and smectite as the main mineral phases of this sample (Horton Jr *et al.*, 2009).

*Sample preparation*

Before the analysis, the CBIS sample was submitted to different treatments in the laboratory. On the one side, coarse and fine powders were prepared to carry out the multi analytical characterization of the sample by simulating the operation conditions imposed by the Analytical Laboratory Drawer (ALD) of the ExoMars rover. Coarse-powder (grains size up to 500 µm) was obtained by crushing a fraction of the solid sample in a sling mill, while fine-powder (grains size below 150 µm) was prepared by crushing the coarse material in an agate mill (McCrone Micronizer Mill) for 12 minutes. On the other side, a thin section was prepared and analyzed to obtain detailed information about the composition of the cementing matrix and the distribution of mineral aggregates. For this purpose, a sample slab was glued to a glass microscope slide using blue stained epoxy resin and polished until reaching a final thickness of 30 µm.

**2.2 Analytical instruments**

*2.2.1 Raman systems*

In this study three different Raman spectrometers were used.

A micro-Raman system equipped with a 633 nm excitation laser was used for the grain by grain study of the coarse powder. The spectrometer, assembled in the laboratory using commercial components, is composed of a Research Electro-Optics LSRP-3501 laser (Helium-Neon), a Kaiser Optical Systems Inc. (KOSI) a HFPH Raman probe, a KOSI Holospec1.8i spectrometer and an Andor DV420A-OE-130 CCD. The system is coupled to a Nikon Eclipse E600 microscope that can focus on the sample with interchangeable long WD objective of 5x, 10x, 20x, 50x and 100x. Raman spectra were collected in a range of 130- 3780 cm$^{-1}$ with a mean spectral resolution of 4 cm$^{-1}$. Data acquisition was performed using the Hologram 4.0 software.

Additional point by point analyses were carried out using the fine powder by a commercial Renishaw inVia Raman micro-spectrometer (Renishaw, UK). The instrument is equipped with a 785nm excitation diode laser, a charge-coupled device detector (Peltier cooled) and is coupled to a microscope with interchangeable long WD objective of 10x, 20x and 50x. Raman spectra were collected in a range of 50-1200 cm$^{-1}$ with a mean spectral resolution of 1 cm$^{-1}$. The Wire 3.2 software (Renishaw) was employed for data acquisition.

As explained in detail by Lopez-Reyes et al. (2013), the RLS ExoMars Simulator is equipped with a BWN-532 excitation laser (B&WTek) emitting at 532 nm with a maximum power of 120 mW and a BTC162 high resolution TE Cooled CCD Array spectrometer (B&WTek). The system is provided with an optical head with a long WD objective of 50x providing a 50 micron spot on the sample. This is coupled to vertical and horizontal positioners that allow emulating the Sample Preparation and Distribution System (SPDS) that will be mounted in the ExoMars rover (line analysis of a powdered sample). Data acquisition was performed using the LabVIEW 2013 software (National Instruments, USA). The main peculiarity of the RLS ExoMars Simulator is its capacity of working in automatic mode, being this one of the main requirements of Raman spectrometers operating on planetary missions. The series of operations and algorithms carried out by the instrument (i.e. autofocus, signal to noise ratio optimization and integration time/number of accumulations selection) to autonomously obtain high quality spectra are described in the work of Lopez-Reyes and Rull Perez (2017). Using the automatic acquisition system, coarse and fine powders were studied by simulating the operational conditions required by the Analytical Laboratory Drawer (ALD) of the ExoMars rover. In addition, the RLS ExoMars Simulator can also work in two-dimensional movements (not only along a line), allowing for the molecular mapping of a surface, performed to obtain detailed information about the molecular composition and mineralogical distribution of the sample. All Raman spectra were collected in a range of 70 - 4200 cm$^{-1}$ with a mean spectral resolution of 6 - 10 cm$^{-1}$.

All Raman data were then treated and interpreted using the IDAT/SpecPro software, which has been specifically developed to receive, decode and manage the data generated by the RLS instrument on Mars. More information about this software can be found in the work of Saiz et al. (2018).

*Other analytical techniques*

The molecular information obtained from Raman spectrometers was complemented by further analytical studies (XRD, NIR, XRF and LIBS), whose some shall be a part of the payload of the Mars2020 and ExoMars2020 rovers.

A X-ray diffractometer (XRD) was employed to analyze the mineralogical composition of the fine-grinded sample. Concretely a Bruker D8 Advance instrument equipped with a CuKα radiation source (40 kV, 40 mA, and wavelength of 1.5418 Å) was used. Diffractograms were gained using a variable slit, a step size of 0.031° 2θ, and a count time of 1.25 s (bulk) in the interval between 2 and 65° 2θ. Mineral identification of randomly oriented whole rock samples was done using the BRUKER DIFFRAC.EVA software and the Powder Diffraction File-2 2002 mineral database (ICDD).

The reflectance spectroscopy in the near-infrared (0.8–4.2 μm) was performed using a Fourier Transform spectrometer in the IR range (PerkinElmer FTIR) under ambient temperature and pressure conditions. The spectral resolution was 4 cm$^{-1}$ and the spectral sampling was 0.35 nm.

The collecting spot size was about 1 mm, thus, representative of all constituents of the fine powder. To calibrate the sample reflectance spectrum, reference spectra were acquired using an Infragold and a Spectralon 99% (Labsphere).

With regards to elemental analysis, the fine-grinded sample was studied using a remote LIBS assembled in the UVa-CSIC-CAB Associated Unit ERICA laboratory. The system consists of the following commercial components: a BIG SKY Ultra CFR Nd:YAG double frequency laser (1 Hz, 45 mJ at 532 nm, 6 ns pulse width), a 70 mm diameter converging lens, a ME-OPT-0007 (Andor) collector and a Mechelle 5000 spectrometer (Andor) operating in the range 200–975 nm. The optical emissions of the plasma pen were collected at an angle of 30 °. The delay between the laser pulse and the acquisition of the emission was 2 ns, while the acquisition time was 8 ns. To perform LIBS analysis, 200 mg of fine grinded sample was pressed under 10 tons (CrushIr,PIKE technologies) for 10 minutes to create an uniform pellet. Each analysis was performed in terrestrial atmospheric conditions and represents the average of 50 laser pulses. Data acquisition was carried out through the Solis software (Andor).

Elemental mappings of thin sectioned samples were carried out using the M4 TORNADO Energy Dispersive X-ray Fluorescence spectrometer (ED-XRF, Bruker Nano GmbH, Germany). This instrument is equipped with a micro-focus side window Rh X-ray tube and polycapillar lenses that allow performing elemental maps with 25 µm of lateral/spatial resolution. A XFlash® silicon drift detector with 30 mm$^2$ sensitive area and energy resolution of 145 eV for MnKα was used for fluorescence radiation detection. In order to improve the detection of the light elements, measurements were acquired under vacuum (20 mbar). The M4 TORNADO software (Bruker Nano GmbH, Germany) was employed for data acquisition and treatment.

To conclude, thin section optical observations were carried out using an Optiphot-pol petrographic microscope (Nikon, Japan). The system is equipped with a rotatory stage, interchangeable objectives of 4x, 10x and 20x, a polarizer filter, an halogen lamp bulb as light source and a digital camera for image collection.

### 3: RESULTS AND DISCUSSION

### 3.1 Coarse and fine powder characterization

The first objective of this work consisted in carrying out a comprehensive characterization of the CBIS sample. To do so, the spectroscopic study performed through LIBS, NIR and Raman systems were complemented and compared to the mineralogical results obtained by XRD.

*3.1.1 LIBS analysis*

The elemental analysis of the fine-powder pellet was carried out in triplicates so as to evaluate its homogeneity. As displayed in Figure 1a, the LIBS spectrum collected from the WH16-014 sample shows a large number of peaks. Most of these signals are due to the presence of iron, which plasma generates several emission lines. Indeed, based on the information provided by the NIST data base, iron provides a total of 6877 atomic lines, mainly located in the UV-Vis range of the spectrum (from 390 to 550 nm, see Figure 1e). Similarly, the element Ca was identified thanks to its main peaks at 393.4 and 396.8 nm (Figure 1d). Calcium provides many signals of high and medium intensity that, together with the iron ones, greatly complicate the detection of additional elements. Despite this limitation, the detection of very intense peaks at 394.4 and 396.1 nm corroborates the presence of aluminum in the CBIS sample (Figure 1d). Similarly, the intense doublets at 588.9-589.5 nm (Figure 1f) and 766-5-769.9 nm (Figure 1g) confirm the detection of sodium and potassium respectively. As shown in Figure 1b, the three emission lines at 279.5 and 280.2 and 285.2 nm (see Figure 1b) proves the presence of Mg, while the sharp signal at 288.1 is provided by Si. As displayed in Figure 1c, the 4 atomic lines at 334.9, 336.1, 337.3 and 338.4 nm are associated with the Ti element. However, the low intensity of these peaks as well as the presence of multiple Fe and Ca signals in this spectral range, can lead to an erroneous interpretation.

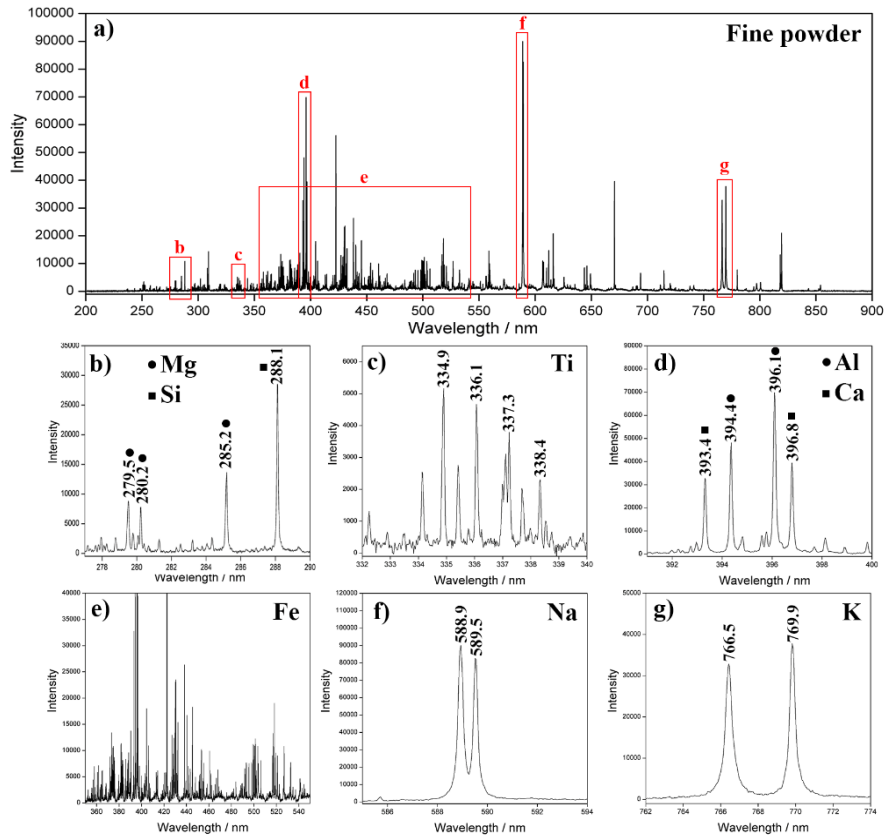

*Figure 1: a) LIBS spectra collected from the fine powder of sample WH16-014. The main emission lines of b) Mg and Si, c) Ti, d) Ca and Al, e) Fe (mainly), f) Na, and g) K, are also displayed.*

*3.1.2 XRD analysis*

The mineralogical analysis of the fine powder sample was performed by XRD. According to the interpretation of the diffractogram displayed in Figure 2, quartz and its high temperature polymorph cristobalite are present as major mineral compounds. The detection of medium/weak diffraction signals proves that the impact breccia includes plagioclase minerals such as sanidine and/or anorthite, as well the presence of minor amounts of feldspars inclusions (probably orthoclase). Finally, further minor peaks indicate the trace presence of clay minerals like illite and possible mixed layered phases. As shown in Table 1, these results fit well with those presented in the work of Horton Jr. et al. (2009).

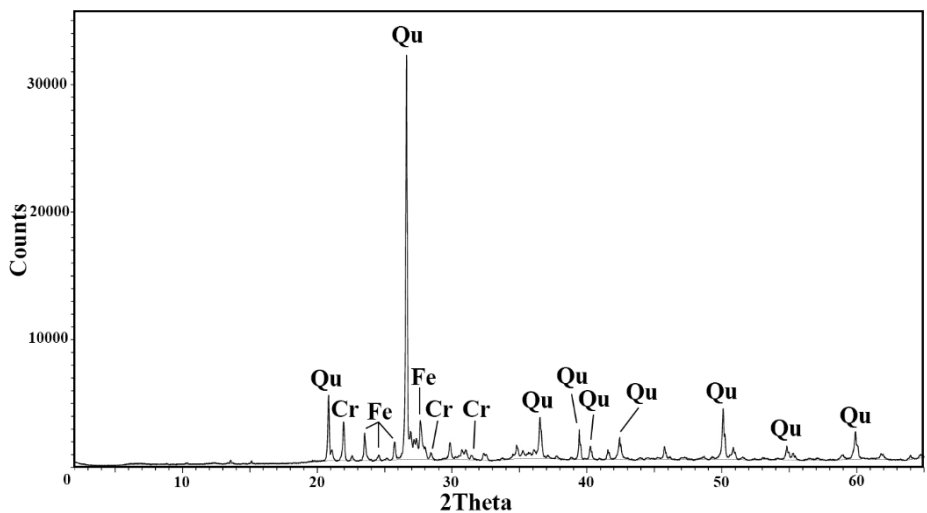

*Figure 2: X-ray diffractogram collected from the fine grinded sample. Peak assignment is represented as follow: Qu= Quartz, Cr= cristobalite, Fe= Feldspars.*

Although X-ray diffractometry is widely employed as routine technique for the mineralogical characterization of geological samples, it is important to underline that the main limitation of this technique consists in the detection limit, which constrains the identification of minor or trace minerals (below 1-5% of concentration depending on the characteristic of the sample and instrumental factors). In the field of planetary exploration, this could represent a great limitation since the detection of minor and trace compounds has a critical impact in defining geological or biological processes that could have occurred on Mars and other extraterrestrial bodies. To overcome this issue, XRD analysis were supported by Raman and NIR spectroscopic studies.

*3.1.3 NIR analysis*

The NIR spectrum obtained from the WH16-014 sample shows several absorption bands. As can be seen in Figure 3, the deep band in the 2.70-3.50 µm region is diagnostic of OH and $H_2O$ stretching in phyllosilicates (combination of bands at 2.76, 2.90 and 3.07 µm) added to the presence of water. The presence of $H_2O$ is also detected near 1.90 µm, while the 1.4 µm is related to OH vibrations. A weak band located at 2.20 µm can be attributed to Al-OH and/or Si-OH, which is in good agreement with the identification of Al-bearing hydrated silicates in the 2.70-3.10 µm range.

Some carbon is detected with the bands at 3.40-3.50 µm associated to organics but some contamination may occur in this spectral region, and the absence of any signature at 3.9 µm is not in favor of the carbonate identification. A broad band between 1.40 and 2.40 µm and a shallow one between 1.00 and 1.30 µm are present but can hardly be attributed to any mineral.

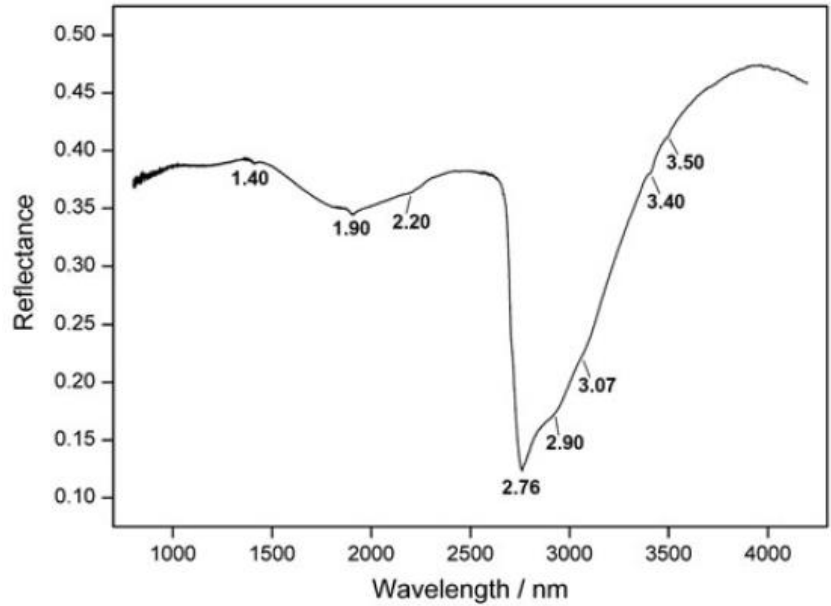

Figure 3: NIR *spectra collected from the fine powder of sample WH16-014.*

*3.1.4 Raman analysis*

*Laboratory system (633 nm excitation laser)*

The laboratory Raman system equipped with a 633 nm emission laser was used to perform a grain by grain study of the coarse powder.

As represented in Figure 4, the interpretation of more than 130 spectra enabled the identification of several compounds that were not detected by XRD and NIR.

Most of the analyzed crystals, visually recognizable for their bright grey color, provided the characteristic spectrum of quartz ($SiO_2$, main peak at 464 $cm^{-1}$). Considering the geological contest, quartz can be part of the pre-impact geology of the target as well as the result of a post-impact diagenesis. However, the first hypothesis is supported by the detection of a slight shift of the main quartz peak towards lower wavelength, which indicates the occurrence of impact-related alteration processes. (the interpretation of this spectral feature is deepened in section 3.2).

Furthermore, as shown in Figure 4a, a very limited number of quartz spectra also displayed a weak additional signal at 419 $cm^{-1}$. Considering that the sample comes from the breccia of an impact crater, this vibrational peak can be interpreted as the evidence of a partial transformation of quartz into coesite ($SiO_2$). Through the detection of coesite, additional information regarding temperatures and pressures reached during the impact can be extrapolated. In fact, the pressure/temperature phase diagram presented by Ford et al. (2004) shows that this polymorph of silicon dioxide is usually formed at temperatures above 800 °C and in a range of pressures ranging from 2 to 10 GPa (Ford *et al.,* 2004).

Furthermore, in about 30% of the quartz spectra collected from this sample, two additional vibrational signals were detected at 230 and 416 $cm^{-1}$ (Figure 4b). These peaks corroborate the presence of cristobalite. According to XRD and Raman studies performed in previous works, the presence of this high temperature polymorph of $SiO_2$ seems to be restricted to the core section between 1402.02 and 1407.49 m of depth (clast-rich impact melt body) (Jackson *et al.*, 2011).

As shown in Figure 4c, the detection of signals at 284 and 514 $cm^{-1}$, together with the right shoulder (474 $cm^{-1}$) of the main peak of quartz, confirms the presence of feldspar crystals in the sample. Even though the detection of the secondary peak at 284 $cm^{-1}$ suggests the presence of a k-feldspars (sanidine or orthoclase) (Freeman *et al.,* 2008), the quality of the obtained spectra does not permit to identify the exact compound within this mineral group.

The intense band at 683 $cm^{-1}$ detected in the same spectrum indicates the presence of ilmenite ($FeTiO_2$), a mineral that has never been detected in the CBIS stratigraphy. Ilmenite, along with rutile ($TiO_2$, Figure 4d, main peaks at 242, 446 and 611 $cm^{-1}$), certifies the presence of titanium-based oxides in the mineralogy of the impact breccia.

The vibrational profile displayed in Figure 4e is assigned to hematite (main peaks at 221, 291, 406, 607, 654 and 1315 $cm^{-1}$), whose presence is associated to alteration processes occurred in the post-impact phase. In fact, the spectra obtained from the analyzed sample differ from standard $Fe_2O_3$ for the additional band at 654 $cm^{-1}$, which is well known to be the consequence of the exposition to high temperatures (de Faria and Lopes, 2007).

Finally, the Raman analysis of the few white crystals visually recognized in the coarse powder provided vibrational spectra with a very intense peak at 989 $cm^{-1}$ along with numerous secondary signals at 460, 617, 630, 648 and 1144 $cm^{-1}$. These spectra perfectly fit with the one provided by the standard of barite ($BaSO_4$). According to the review of Hanor (2000), barite crystallization can be associated to 1) biologically mediated precipitation, 2) hydrothermal activity 3) submarine discharge of fluids from continental margins and 4) drilling mud contamination. Considering that the CBIS is a wet-target crater and that the process of impact breccias cooling last several thousands of years (Osinski *et al.,* 2005), it can be deduced that the formation of barite was due to the development of a post-impact hydrothermal system (Sanford, 2005). Furthermore, the hypothesis relating the detection of this compound to a contamination occurred during drilling is excluded by the fact that that barite minerals were observed in the inner part of matrix rather than in the external surface of the drill core.

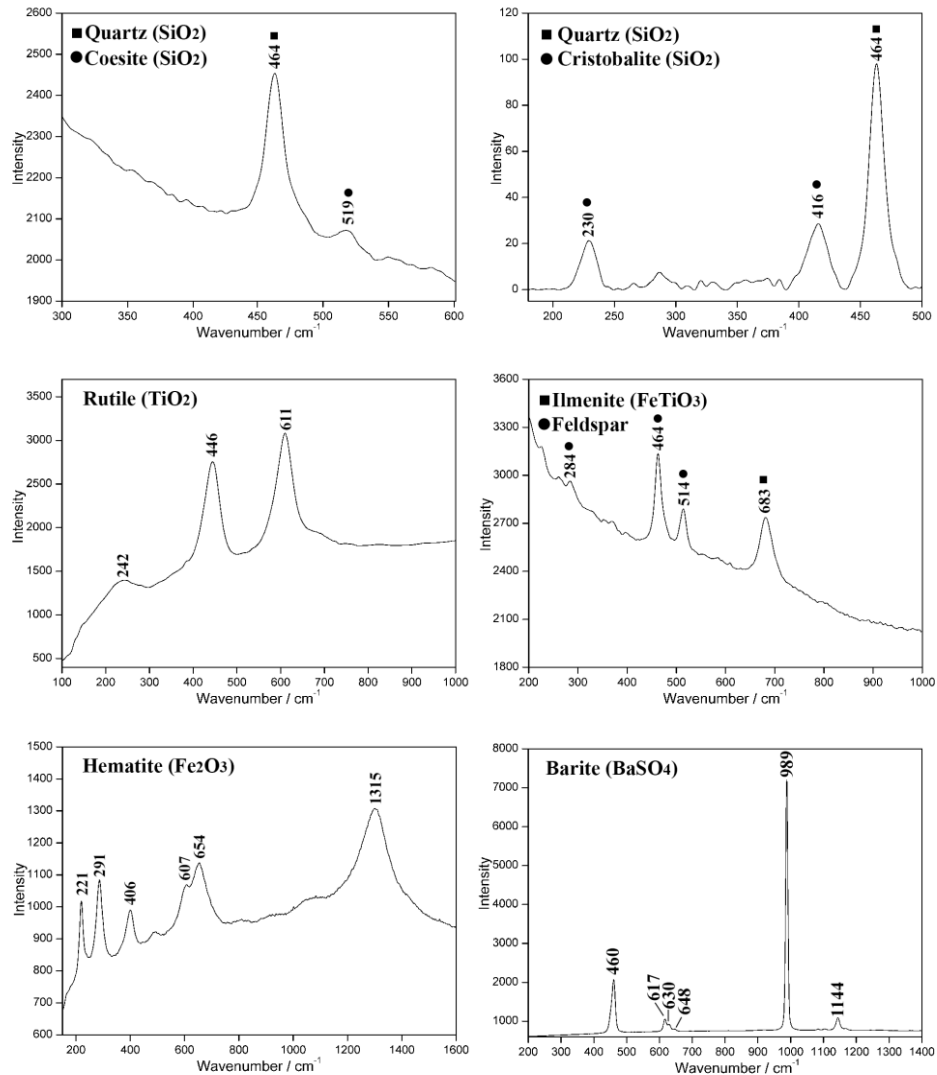

*Figure 4: Raman spectra collected from the coarse powder of sample WH16-014 by means of the laboratory system (633 nm excitation laser).*

*RLS ExoMars Simulator (532nm excitation laser)*

The results obtained through spectroscopic and diffractometric laboratory systems were then compared to the molecular data provided by the RLS ExoMars Simulator. As mentioned before, the purpose of these analysis was to evaluate whether the simulator is capable of confirming the mineralogical composition of the sample by emulating the analytical procedure that will be employed by the ExoMars rover.

To achieve this aim, the sample holder of the SPDS was filled with coarse powder and Raman analysis were carried out in a defined grid of points. By using the same algorithms that will be employed by the Raman spectrometer on Mars, the signal to noise ratio, the autofocus, and the optimal ratio between number of accumulation and time of acquisition were automatically calculated for each spectra.

Through the RLS ExoMars Simulator, the main minerals composing the WH16-014 sample were detected, such as quartz, coesite, rutile, k-feldspars, hematite and barite. Furthermore, as displayed in Figure 5, the identification of Raman spectra showing vibrational peaks at 297 and 1090 cm$^{-1}$ proved the additional presence of siderite (FeCO$_3$), an iron carbonate that was not identified through the use of laboratory Raman systems. As for the case of barite, siderite crystallization can take place under hydrothermal conditions among others (Treiman *et al.,* 2002).

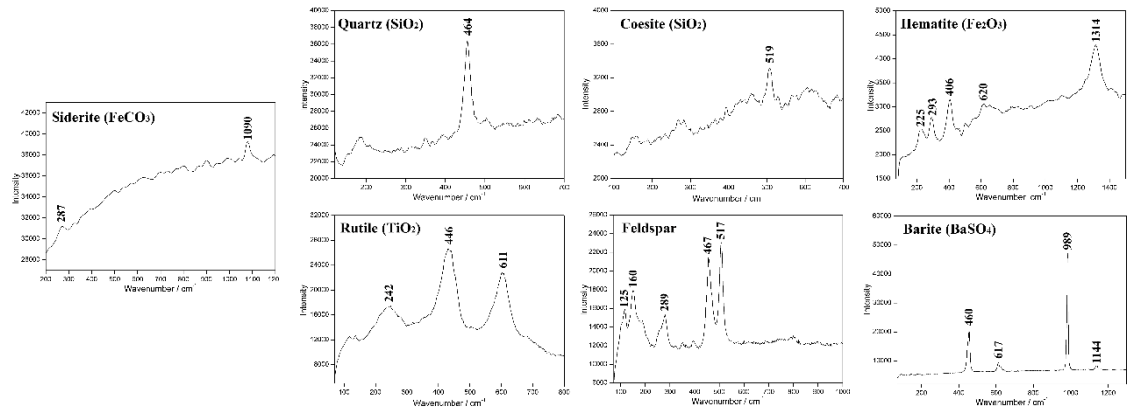

Figure 5: *Raman spectra collected from the coarse powder of sample WH16-014 by means of the RLS ExoMars Simulator (532 nm excitation laser).*

With regards to quartz spectra, the analysis performed with the RLS ExoMars Simulator also showed the shift of the main peak towards lower wavelengths. This aspect is discussed in section 3.2.

Concerning the detection of barite and siderite, it is important to emphasize that these compounds are considered as possible mineral indicators of aqueous environments in Mars (Treiman *et al.,* 2002; Burt *et al.,* 2004). In this sense, the detection of hydrothermal compound through the RLS ExoMars Simulator has critical relevance in the field of Mars exploration. Indeed, these results suggest that the Raman system that will land on Mars as part of the payload of the ExoMars rover has the potential to analytically confirm the wet-target origin of Martian craters, and consequently the past presence of water on the planet surface.

*3.1.5 Result comparison*

Table 1 summarizes and compares the analytical information provided by the instruments used in this work with those available in bibliography.

As explained above, the XRD results obtained in this study agree with those presented by Horton Jr. et al. (2009). The only difference resides in the identification of the mica, since the work of 2009 indicates the presence of smectite while this study suggests the presence of illite and other possible mixed phases. The NIR system confirmed the presence of such Al-OH hydrated phase, while the other minerals detected by Raman and XRD were not identified since they are largely featureless in the IR range. Regarding Raman analysis, both the laboratory system and the RLS

ExoMars Simulator were able to detect minor and/or trace compounds that were not identified by XRD or NIR, proving a higher complexity of the mineralogical composition of the sample.

*Table 1: Comparison of mineralogical data collected from sample WH16-014 from different diffractographic and spectrographic techniques.*

| XRD (Horton Jr. et al (2009)) | XRD (This work) | NIR (This work) | Raman - Laboratory system (this work) | Raman – RLS ExoMars Simulator (This work) |
|---|---|---|---|---|
| quartz | quartz | | quartz | quartz |
| cristobalite | cristobalite | | cristobalite | |
| hydrated silicate (smectite) | hydrated silicate (illite) | Al-bearing hydrated silicate (illite-muscovite) | | |
| feldspar (sanidine) | feldspar (sanidine/anorthite and orthoclase) | | k-feldspar | k-feldspar |
| | | carbon (contamination?) | | |
| | | | coesite | coesite |
| | | | rutile | rutile |
| | | | hematite | |
| | | | ilmenite | |
| | | | barite | barite |
| | | | | siderite |

**3.2 Shock metamorphisms**

As explained in the previous section, Raman analyses detected a slight displacement of the predominant band of quartz (464 cm$^{-1}$) towards lower wavelengths, proving the shock-metamorphism suffered by $SiO_2$ grains due to pressures reached during impact (McMillan *et al.*, 1992).

Fritz et al. (2011) suggest that the main band of quartz is produced by the vibration of oxygen atoms connecting two silica tetrahedra and its shift towards lower wavelengths is due to the increase of the angle spanning two tetrahedra. Furthermore, the irregularities of the crystal structure (bond lengths and bond angles) also leads to an increase of the main band width.

The degree of peak displacement and broadening is proportional to the strength of the impact. Therefore, the Raman characterization of this phenomenon is of primary importance in the field of planetary exploration, representing the analytically clue that confirms the impact origin of a crater, provided quartz is present in the target.

*3.2.1 Analysis of coarse powder*

A detailed grain by grain analysis of the coarse powder was carried out to evaluate the degree of the amorphization phenomenon. 100 quartz grains were first analyzed by means of the laboratory Raman system (633 excitation laser) and the main peak of each spectrum was integrated to calculate the exact band position and the Full Width at Half Maximum (FWHM) values.

As represented in Figure 6, the values obtained from the CBIS sample were compared to those provided by 4 sample of polycrystalline quartz crystals proceeding from different geological contexts. The image clearly illustrates that most of the quartz fragments included in the WH16-014 sample show shock amorphization effects. Indeed, several main peaks are displaced from 0.5 to 2.5 cm$^{-1}$ with respect to the standards, with highs of up to 4.0 cm$^{-1}$. The graph also shows a clear tendency according to which the FWHM value increases by increasing the peak displacement. Comparing the obtained values with the one proposed in the work of McMillan *et al.* (2011), it can be deduced that the sample was subjected to pressures equal to or greater than 25.8 GPa.

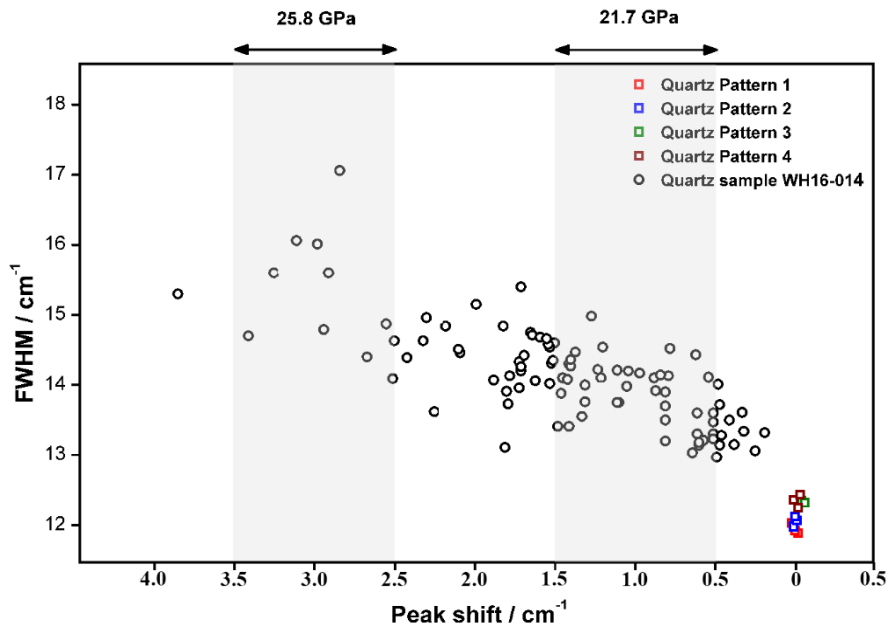

*Figure 6: Graph representing the displacement and width values of quartz Raman peaks obtained from the coarse grinded sample. Gray areas indicate the peak shifting values (with a confidence interval of 1cm$^{-1}$) measured by McMillan et al. (1992) at the shock pressures of 21.7 GPa and 25.8 GPa respectively.*

Analyzing in detail the most characteristic spectra of the sample (Figure SM1), it was possible to observe that, by increasing the shift of the main peak of quartz, the band became broader and asymmetrical. Thus, peaks were integrated to extrapolate the band parameters necessary to calculate and quantify the degree of amorphization. In the case of non-shocked quartz grains, the peak maximum is located at 463.7 cm$^{-1}$ and the signal shape is symmetrical. Indeed, the half width value of the left part (HWL= 7.00 cm$^{-1}$) of the peak is equal to the right one (HWR= 7.00 cm$^{-1}$). In the second case (slightly shocked quartz), the shift of the main peak (462.9 cm$^{-1}$) is accompanied by a slight asymmetry, being the HWR value (7.40 cm$^{-1}$) higher than the HWL one (7.20 cm$^{-1}$). Finally, the HWR value of the strongly amorphized quartz (461.3 cm$^{-1}$) increases considerably compared to the previous spectra (8.40cm$^{-1}$), while the HWL remains unchanged (7.20 cm$^{-1}$). This phenomenon is related to the loss of coherence of the crystal lattice of shocked quartz. Indeed, the Raman scattering resulting from the vibration of amorphized crystals having high and low periodicity zones causes both the broadening and the asymmetry of the characteristic main peak.

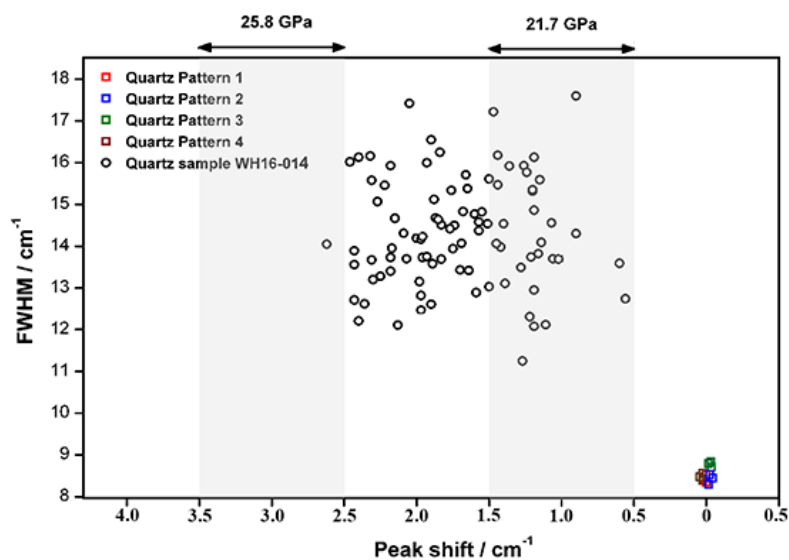

*Figure SM1: Calculation of the main band parameter from characteristic spectra of a) no shocked, b) slightly shocked, c) strongly shocked quartz.*

### 3.2.2 Analysis of fine powder

When Raman analysis are performed on fine powders, unwanted side effects such as background increase and peak widening can take place, causing negative repercussions on the detection capabilities of the spectrometer (Rull *et al.,* 2017). Considering that the particle size distribution of samples crushed by the SPDS of the ExoMars rover is comprised between 2 to 700 µm, a point by point study of the fine powders (grain size below 150 µm) was performed to evaluate if the mentioned side effects can hinder the Raman detection of quartz amorphization.

Fine powder was first analyzed by the laboratory InVia Raman system that, having a spectral resolution of 1 cm$^{-1}$, ensures the best performance in terms of peak shift detection. The results were then compared to those provided by the RLS ExoMars Simulator, which allows to realistically estimate the extent of the analytical results that could be obtained by RLS on Mars. In both cases, analyses were carried out in automatic mode by selecting a series of spots within a grid of 12 x 12 points (144 spectra).

The spectra obtained using the laboratory instrument detected amorphous quartz in most of the analyzed spots. In this case, however, most of the values were located in the range of peak shift values from 1.0 to 2.5 cm$^{-1}$ (Figure SM2). The grouping of shift values is due to the fact that, using a 50x objective, the spot of analysis is larger than the size of single grains. For this reason each spectrum has to be interpreted as the sum of the vibrational profiles provided by several quartz crystals with different degrees of amorphization.

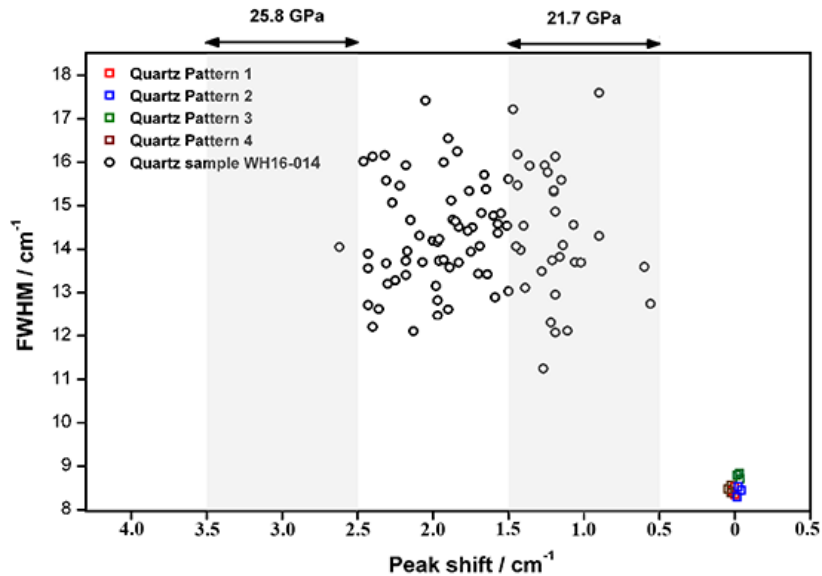

*Figure SM2: Graph representing the displacement and width values of quartz Raman peaks obtained from the fine grinded sample using the laboratory InVia system.*

Analysis with the RLS ExoMars Simulator were carried out in automatic mode using the algorithms for autofocus, intelligent reduction of the spectrum fluorescence and noise signal, and optimization of the measurement conditions (signal integration time and number of accumulations) that will be employed on Mars.

The results (Figure 7) clearly demonstrated that this system is capable of identifying the shock amorphization of quartz. In spite of the background increase and peak widening produced by the fine powder, the shift of the quartz peak was clearly detected. This result has important repercussions in the field of planetary exploration, indicating that the RLS system has the capability of analytically demonstrate the impact origin of Martian craters.

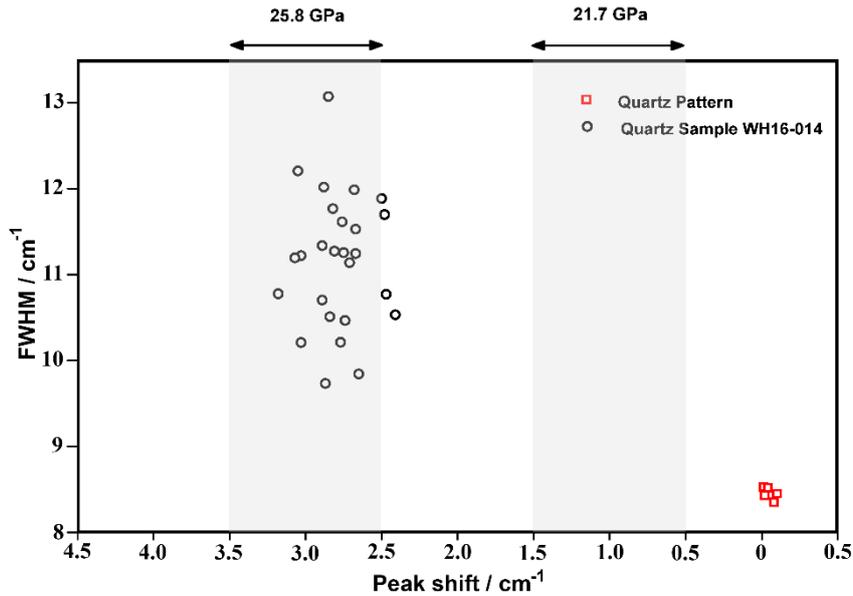

*Figure 7: Graph representing the displacement and width values of quartz Raman peaks obtained from the fine grinded sample using the RLS ExoMars Simulator.*

### 3.3 Thin section characterization

In the last part of this study, additional analyses were made on a thin section to obtain information about the spatial distribution of the detected crystalline compounds and the composition of the cementing matrix.

*3.3.1 Petrographic microscope*

According to the petrographic observation of the thin section, the CBIS sample presents the typical complex features of impact melt breccias. It contains large amounts of melt as well as dispersed grains of quartz. Feldspars and rock fragments (sedimentary) are also found floating in this matrix. In several areas of the thin section, crystal particles seem to be covered by rims of clay minerals.

In some of the detected quartz crystals, the presence of thin and isotropic lamellae was clearly observed (Figure SM3). This set of planar deformation features (PDFs) is widely acknowledged as the most convincing "low" pressure indicator of impact-related amorphization of crystals. However, just a limited fraction of the observed grains contains PDFs, indicating that the majority of quartz crystals are not shocked or weakly shocked (to generate visible planar fractures and planar deformation features, pressures above 1 GPa are needed).

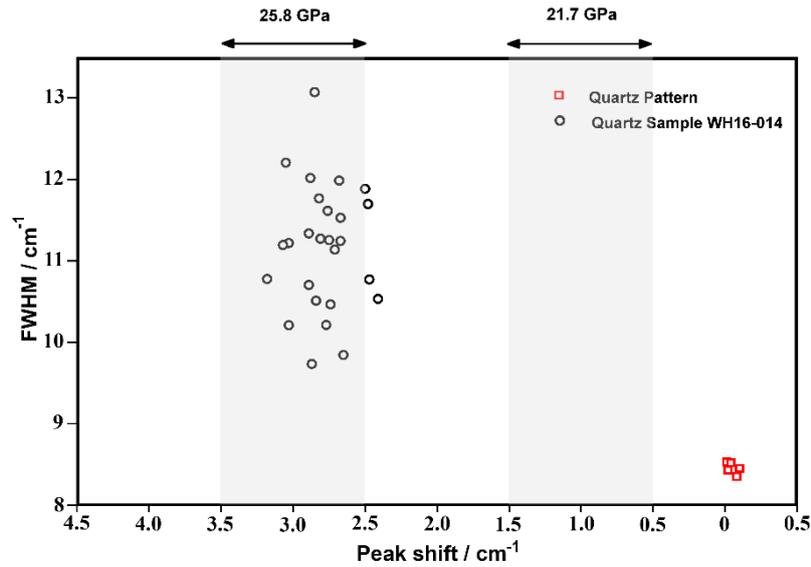

*Figure SM3: Optical image of the thin section obtained through the use of the Optiphot-pol petrographic microscope. The main quartz crystal located in the middle of the figure displays 2 sets of planar deformation features, indicative of high shock pressures.*

*3.3.2 XRF analysis (elemental mapping):*

The elemental distribution in the maps provided by the XRF instrument (Figure 8) confirms the great heterogeneity of the sample. As suggested by the observation of the thin section under a petrographic microscope, the numerous areas rich in Si suggest the extensive presence of quartz crystals of different size and shape. On the other hand, the overlapping distribution of Si, Al, Ca and K in several areas of the thin section should be interpreted as inclusions of feldspars or other silicate minerals. The identification of small inclusions of pure Fe fits with the detection of hematite by Raman spectroscopy. To conclude, small spots presenting a considerable concentration of Ti and S were detected. According to the molecular results previously described, these inclusions can probably be associated with titanium oxides (rutile and ilmenite) and barium sulfate (barite) respectively.

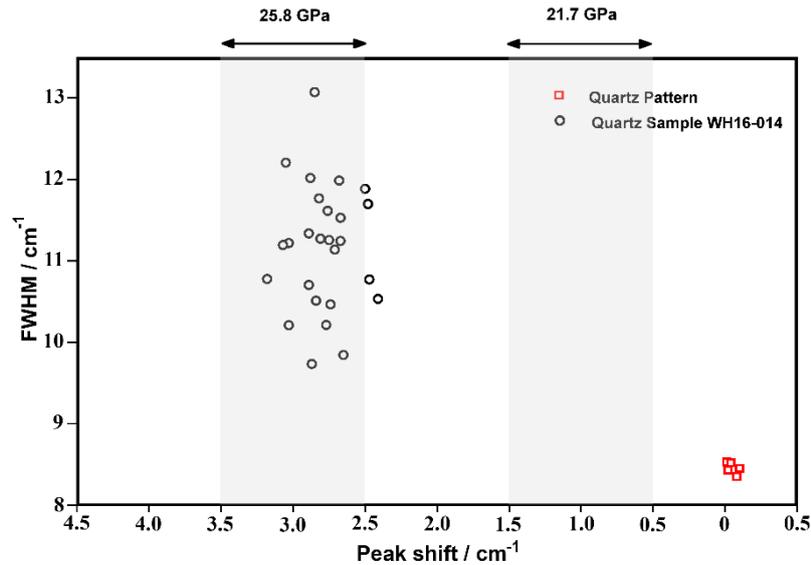

*Figure 8: Elemental mapping of the thin section prepared from sample WH16-014.*

*3.3.3 Raman analysis (molecular mapping)*

The elemental data summarized in section 3.3.2 was used to select the most interesting area for molecular mapping. This task was performed using the RLS ExoMars Simulator to 1) evaluate if the developed algorithms are capable of automatically adapting the analytical parameters so as to obtain optimal spectra from the sample minerals and 2) explore the scientific capabilities of the RLS instrument beyond the analytical requirements or possibilities of the ExoMars mission.

According to XRF results, the selected area (1200 x 1200 µm) should present several crystals of quartz and titanium oxides. To confirm this hypothesis with molecular data, the area was automatically scanned by the RLS ExoMars Simulator selecting 25 µm of lateral/spatial resolution. After completing the automatic collection of the 2500 Raman spectra, MatLab routines were used to elaborate the acquired data and create the molecular distribution images that are represented in Figure 9. These analyses were performed by targeting the spectral regions of the expected materials (rutile and quartz), on automatically baseline-removed spectra, using the same routines implemented on the IDAT/SpectPro software. The molecular map displayed in Figure 9 clearly highlights the presence of several quartz crystals (blue color) together with minor rutile inclusions (red color). With regards to Figure 9c, the intensity of the green spots is directly correlated to the intensity of the wavelength shift of quartz. From this map it is inferred that evidences of shock-metamorphism are equally distributed in all quartz crystals rather than recorded on single isolated grains. The graphical representation of the quartz shift and FWHM values of each quartz spectrum collected from the thin section displays a tendency similar to the one shown in Figure 6.

Similarly to quartz, the detected rutile crystals also showed the spectroscopic evidence of shock metamorphism (slight amorphization and shift of the main peaks at 446 and 611 cm$^{-1}$). This data proves that rutile was present in this stratigraphic unit before the impact occurred.

Finally, the spots of analysis falling between crystals did not provide any clear Raman signal, suggesting the presence of amorphous material. This result agrees with microscope observations and analytical data from previous works, which identified melt rock as the cementing matrix of this geological unit (Rodriguez *et al.,* 2005)

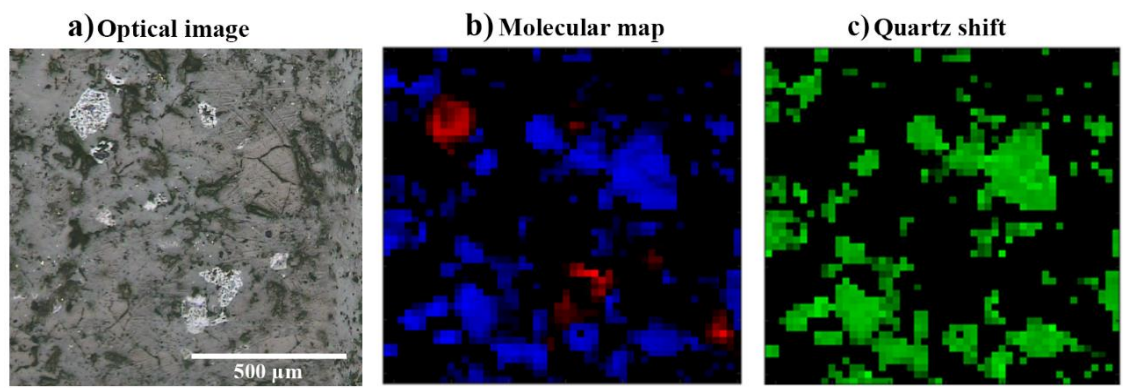

*Figure 9: Molecular maps of the thin section obtained by using the RLS ExoMars Simulator. a) optical image of the selected area, b) spatial distribution of the detected compounds (the intensity of the colors is directly proportional to the intensity of the main Raman peaks), and c) degree of quartz shifting (the intensity of the green color is directly proportional to the shifting of the main quartz peak).*

**4 CONCLUSIONS**

The first objective of the present work consisted in obtaining a multi analytical characterization of the CBIS analogue sample, which will be included in the Planetary Terrestrial Analogue Library (PTAL) database. Even though the petrographic study of coarse-grained rocks is generally performed by supporting optical thin section observations with XRD analysis, the results presented in this paper prove that additional information can be obtained by performing complementary spectroscopic analysis.

In this case of study, Raman spectroscopy enabled the identification of minor compounds that were not detected by XRD, such as coesite, rutile, hematite, ilmenite, barite and siderite. The environmental and geological conditions necessary for the crystallization of these minerals, contributes to deepen the knowledge about the post-impact geological processes occurred in the CBIS. For instance, the detection of secondary hydrothermal products such as coesite and barite is particularly interesting since provides the spectroscopic evidence confirming the occurrence of mineralization processes associated to the presence of water in the bolide-target and triggered by the high temperatures generated by the impact.

The molecular data obtained from the use of the RLS ExoMars Simulator were compared to those provided by laboratory Raman systems aiming at extrapolating valuable insight about the scientific capabilities of the RLS instrument on Mars. In this way, most of the mineral phases composing the sample were effectively identified. Among them, the detection of shocked quartz, barite and siderite could have a strong scientific relevance in the field of planetary exploration. On one hand, the detection of quartz shifting confirms that this instrument is capable of detecting mineral amorphizations induced by high pressure and/or temperatures, which help to confirm the impact origin of a crater. On the other hand, the detection of barite and siderite represents the analytical data that could confirm the presence of water at the impact site.

Considering that the RLS ExoMars Simulator represents a reliable tool to effectively emulate and predict the scientific capabilities of the RLS instrument, the results presented in this work suggest that the ExoMars/Raman system that will be deployed on Mars could have the capability to analytically discriminate dry and wet-target craters.

Beyond planetary exploration issues, the analysis performed on the thin section proves that the RLS ExoMars Simulator is capable of producing high quality molecular distribution maps. Optimal results are guaranteed by the algorithms developed for the automatic operation of the Raman system, which adapts the acquisition parameters according to the characteristics of the spot under analysis. This unique feature represents a great advantage over most commercial systems, which use constant acquisition parameters throughout the scanned surface.

**Acknowledgments**

This project is financed through the European Research Council in the H2020- COMPET-2015 programme (grant 687302) and the Ministry of Economy and Competitiveness (MINECO, grants ESP2014-56138-C3-2-R and ESP2017-87690-C3-1-R). The authors would like to thank W.R. Horton Jr. (U.S. Geological Survey) for providing analogue material analysed in this work (WH16-014), and C. Marcilly and T. Naido (University of Oslo) for carrying out sample preparation and XRD analysis.**References**

Barlow, N. and Perez, C.B. (2003) Martian impact crater ejecta morphologies as indicators of the distribution of subsurface volatiles. *Journal of Geophysical Research: Planets* 108:N08.

Bartosova, K., Gier, S., Horton Jr, J.W., Koeberl, C., Mader, D. and Dypvik, H. (2010) Petrography, mineralogy, and geochemistry of deep gravelly sands in the Eyreville B core, Chesapeake Bay impact structure. *Meteoritics & Planetary Science* 45:1021-1052.

Bost, N., Ramboz, C., LeBreton, N., Foucher, F., Lopez-Reyes, G., De Angelis, S., Josset, M., Venegas, G., Sanz-Arranz, A., Rull, F., Medina, J., Josset, J.L., Souchon, A., Ammannito, E., De Sanctis, M.C., Di Iorio, T., Carli, C., Vago, J.L. and Westall, F. (2015) Testing the ability of the ExoMars 2018 payload to document geological context and potential habitability on Mars. *Planetary and Space Science* 108:87-97.

Burt, D.M., Kirkland, L.E. and Adams, P.M. (2004) Barite and Celestine Detection in the Thermal Infrared- Possible Application to Determination of Aqueous Environments on Mars. Proceeding from the Lunar and Planetary Science XXXV conference, Houston (EEUU).

Collins G.S. and Wünnemann K. (2005) How big was the Chesapeake Bay impact? Insight from numerical modeling. *Geological Society of America* 33:925-928.

Debus, A., Bacher, M., Ball, A., Barcos, O., Bethge, B., Gaubert, F., Haldemann, A., Kminek, G., Lindner, R., Pacros, A., Rohr, T., Trautner, R. and Vago J. (2010) Exomars 2010 Rover Pasteur